\documentclass[twocolumn,showpacs,preprintnumbers,amsmath,amssymb]{revtex4}

\usepackage{graphicx}
\usepackage{dcolumn}
\usepackage{bm}

\newcommand{\vett}[1]{\mathbf{#1}}
\newcommand{\vir}{\mathcal{V}}

\newcommand{\dqj}{ {\dot q}^{(j)} }
\newcommand{\vqj}{ {\vett q}^{(j)} }
\newcommand{\eff}{\textrm{eff}}
\newcommand{\ret}{\textrm{ret}}

\begin{document}

\preprint{APS/123-QED}

\title{Far Fields and Dark Matter}

\author{Andrea Carati}
 \email{carati@mat.unimi.it}
\affiliation{%
Department of Mathematics, University  of  Milano\\ 
Via Saldini 50, 20133 Milano, Italy
}%

\author{Sergio L. Cacciatori}
 \email{sergio.cacciatori@uninsubria.it}
\affiliation{ Department of Physical and Mathematical Sciences, 
              Insubria University \\ 
              Via Valleggio 11, I-22100 Como, Italy 
}%

\author{Luigi Galgani}%
 \email{galgani@mat.unimi.it}
\affiliation{%
Department~of Mathematics, University  of  Milano\\ 
Via Saldini 50, 20133 Milano, Italy
}%

\date{\today}

\begin{abstract}
We give an estimate of the gravitational field of force exerted on a
test particle by the far galaxies, in the frame of the weak field 
approximation. In virtue of Hubble's law, the action of the far matter
turns out to be non negligible, and  even the dominant one.
A nonvanishing contribution is obtained only if the discrete and
fractal nature of the matter distribution  is taken into account. The
force per unit mass acting on a test particle is found to be of the
order of $0.2\, cH_0$, where $c$ is the speed of light and $H_0$ the
present value of Hubble's constant.
\end{abstract}

\maketitle
The main idea underlying the present paper  comes from a critical reading
of the first work in which the existence of dark matter was proposed,
namely the work of Zwicky \cite{zwi} in which the virial 
theorem was applied to the Coma
cluster. The point concerns the role of the forces due to the external
matter. Let us recall that in the first use of the virial theorem, namely
that of Clausius for a gas enclosed in a vessel,  the main contribution
to the virial was coming from the external forces (the
pressure of the walls),  because the contribution of the internal
forces was either vanishing (for a perfect gas) or negligible.   
In the case of Coma,  the contribution of the
internal forces was still found by Zwicky to be negligible, but 
the contribution of the external galaxies 
 was neither taken into account,  nor 
even mentioned at all, perhaps because one could not conceive how it
may act as a pressure.  The missing contribution to the virial
was thus attributed to some hypothetical interior dark matter. 
In the present letter we  show how it is conceivable that
 the external matter (actually, the
far one) may produce  a 
pressure, and we also give a rough estimate of its 
contribution to the virial, which appears to be  in  rather good
agreement with  the observations.

The first point we make is that, as the  forces on a test particle depend not
only on the positions of the galaxies, but also on their
velocities and accelerations,  due to Hubble's law the
dominating contribution comes from the far matter. This comes about
as follows.
From the point of view of general relativity, in the 
weak--field approximation the problem of estimating the force (per
unit mass) on a
test particle   amounts
to writing down the equations for the geodesic motion when the  metric
tensor $g_{\mu \nu}$ is a solution
of the Einstein equation with the external galaxies as sources.
Writing  the metric tensor as a perturbation of
the Lorentzian  background $\eta_{\mu \nu}$, namely, 
as $g_{\mu \nu}=\eta_{\mu \nu} +h_{\mu \nu}$, the perturbation  
$h_{\mu\nu}$ turns out to be a solution of the 
wave equation, so that its components are analogous to the familiar
 retarded potentials of electrodynamics (although relevant differences
 exist  between the two cases, as particularly emphasized by Zeldovich
 and Novikov \cite{zeldovich} ).
In fact one finds
\begin{equation}
\label{campo}
h_{\mu\nu}=\frac {-2\pi G}{c^4 }\, \, \sum_j  M_j \frac 1{\gamma_j}\,
\left. \frac {2\dqj_\mu 
\dqj_\nu -c^2\eta_{\mu\nu}} {|\vett x-\vqj|}\right|_{t=t_{\ret}} \ ,
\end{equation}
where $G$ is the gravitational constant, 
while $M_j$, $\vett q^{(j)}$  and $\gamma_j$, $\ j=1,\ldots, N$,  are
the mass, the position vector and the  Lorentz factor of the $j$--th
source galaxy,   the dot denoting  derivative with respect
to proper time along the worldline of the source.

As   $h_{\mu\nu}$ depends 
on each  source not only through its position, but  also through its velocity,
 the latter has to be assigned in order that the model 
be defined.
To this end  we use Hubble's law as a phenomenological
prescription. Making reference to a local chart with
Lorentzian coordinates  having as
origin the center of mass of the considered localized system (the Coma
cluster), for the velocities $\vett q_j\equiv \vett q^{(j)}$ of the
external galaxies we thus require
\begin{equation}\label{hubble}
\dot {\vett q}_j=  \gamma_j^{-1}H_0 \vett q_j\ , 
\quad j=1,\ldots, N 
\end{equation}
(the dot  denoting now  derivative with respect to the background
 Lorentzian time). Here,  $H_0$ is
the Hubble constant which, in our extremely simplified model, we take
fixed at its  present value. 

Notice that the Hubble  assumption (\ref{hubble}) 
has an essential  impact on the size of the gravitational
field of force. Indeed such a force contains  a term  
(decreasing as $1/r^2$) proportional to the velocity
of the source, and a term  (decreasing as $1/r$) proportional to the 
acceleration of the source. Thus, estimating the acceleration through
Hubble's law,   the latter  term
actually doesn't depend on distance at all, so that
 the far matter is found to give the dominant contribution 
 to the gravitational field of force. This situation is reminiscent of
 the way in which Mach's principle was dealt with in
\cite{einstein} (see page 102). The main difference  being
 that in such a case,
 lacking Hubble's law, the velocities of the sources were 
neglected. Consequently,
 only the Newtonian, fast decaying, potential was considered, so that  
only the near matter, and not the far one, appeared to play a role.

So we address our attention   to the dominating term,   proportional 
to the acceleration of the source. Such a term, which we denote by
$\vett f$,  has the form
\begin{equation}\label{vettore}
\vett f= \frac {4G H_0^2\, M}{c^2}\  \vett u\ ,\quad \quad
\vett u(N)= \sum_{j=1}^N \frac{\vett q_j}{|\vett q_j|}\ ,
\end{equation}
where the masses of the galaxies were all put equal to a common
value $M$, and the Lorentz factors $\gamma_j$ were put
equal to 1, for the reasons to be illustrated later.
Notice the extremely simple nature of this force per unit
mass (or acceleration). 
Apart from a multiplicative factor, such a force is just the sum of
all the unit vectors pointing to each of the external galaxies.
Actually, our attention was addressed to   
the component of such a force $\vett f$ along a  given
direction. Such a component will be simply
denoted by $f$, and   the corresponding component of $\vett u$
by $u$.

Having determined the quantity of interest ($f$ or $u$), we come  now to our
second main point, namely the problem of how to describe the distribution
of the external galaxies.
 It is immediately seen that $u$ exactly vanishes (at any
 point)  if the external matter is 
described  as  a continuous medium  with a spherically symmetric  density.
This should be  expected,  in agreement with  Birkhoff's
theorem.
So we take  a different  point of view, 
 analogous to  the  probabilistic  one introduced by
Chandrasekhar and von Neumann (see the review \cite{chandra}) in the problem
of estimating the  vector sum of the Newtonian forces
exerted  on a star   by  the surrounding ones. 
In such an approach, the external sources are  conceived
as point particles, and the place of the  matter density of the
continuum case  is now taken
by a probability density for the position of a galaxy. 
From such a point of view, the previous result (the vanishing of $u$
for a spherically symmetric matter density), now reads as the
vanishing of the mean value of $u$ for a spherically symmetric
probability density of the position of a galaxy.

We thus come  to an estimate of the variance of the  force  $f$ (or of $u$). It
will be seen that the  result depends on the further 
assumptions one introduces concerning  the spatial distribution of
 the external galaxies.  
Assume first that the positions $\vett q_j$ 
 of the $N$ galaxies are independent   random
variables, uniformly distributed with respect to the
Lebesgue measure. Then  the sum defining $u$ is found to grow 
as $\sqrt N$. This  indeed is just a consequence
of the central limit theorem, because  in such a case $u$
is the sum of $N$ independent identically distributed  random
variables which turn out to have  zero mean  and  a finite variance. 
By the way, such a result is the  analogue
 of that obtained by Chandrasekhar and von Neumann for the case of
Newtonian forces,  the only difference being  that in their case 
the variance is infinite (due to the divergence of Newton's 
force at zero distance, a property which plays  no role  in our case).
  For what concerns the corresponding  estimate of the virial, one easily sees
that with the present assumption the estimate
 is by far too small to account for the 
observations, just because the
 considered sum behaves as $\sqrt N$ rather than as $N$ (see later).  

So we modify the previous assumption, and consider the case in which
the probability density is fractal \cite{mandel} (see also
 \cite{sylos}).
This means first of all that
  the positions of the galaxies are no more independently
distributed, so that $u$ is
no more constrained to grow as $\sqrt N$, and can instead have a
faster growth, as required by the observations.
However, the analytical
computation of the probability distribution of the field of force 
 now becomes 
 a  quite nontrivial task, with respect  to the much simpler case
 considered by Chandrasekhar and von Neumann.
So we are forced, at least  provisionally, to   
investigate the  problem by numerical methods. 
\begin{figure}
\includegraphics[width=3.5 in]{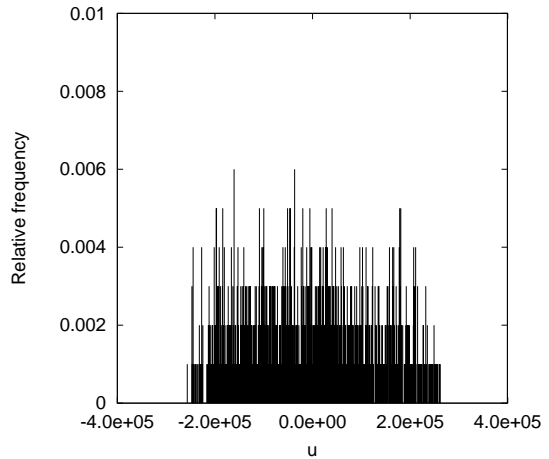}
\caption{\label{fig1} Histogram of the random variable
  $u$, which is proportional to the force per unit mass due to $N$
  external galaxies. The histogram was computed with  $10,000$ 
  samples of $N= 512,000$ galaxies, by counting the fraction  
  of times the value $u$ belongs  to a given interval of width $\sqrt
  N$.
}
\end{figure}

We proceeded as follows.
In order to estimate (at the origin of the
coordinates) the sum defining $u$, the positions of the $N$ 
galaxies are extracted (with the method described
 in \cite{mandel}) in such a way  that the
mass distribution has a fractal dimension, precisely
the fractal dimension $2$. 
 The corresponding histogram, obtained through 10,000
samples of configurations of the galaxies, 
is shown  in  Fig.~\ref{fig1} for $N=512,000$.

We then study the dependence of $u$ on  the number $N$ of external
galaxies, which was made to vary
in the range   $1000\le N\le 512,000$, 
with the density  kept constant. 
This means that the positions of the $N$ points
were  taken to lie inside a cutoff
sphere whose volume was made to increase as $N$. For the values of $N$
investigated, the corresponding radius turns out to be so small with
respect to the present horizon, that the Lorentz factors $\gamma$
could altogether be put equal to $1$ (as was previously assumed), 
and more in general  the special
relativistic character of our model was actually justified.  

\begin{figure}
\includegraphics[width= 3.5 in]{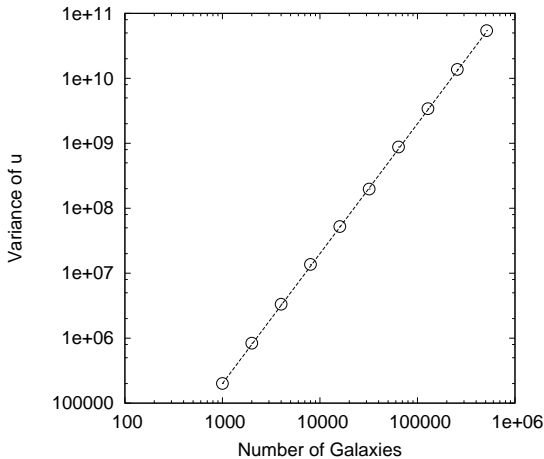}
\caption{\label{fig2} The variance $\sigma^2_u$ of $u$ versus the
  number $N$ of galaxies in log--log scale. The dashed line is the curve
  $\sigma_u^2=0.2\ N^2$.
}
\end{figure}

The mean of $u$ turns out to
practically vanish  for all $N$, while its  variance $\sigma^2_u$ is found to
grow as  $N^2$ (actually, as $0.2\, N^2$),  rather than as $N$, as
occurs  in  the uniform
case. This is shown in Fig.~\ref{fig2}. We thus
conclude that the standard deviation $\sigma _f$ of  the component of
the force per unit mass along a direction is proportional to $N$,
being given  by
\begin{equation}\label{sigmaf}
\sigma_f\simeq \sqrt{0.2}\  \frac {4GH_0^2}{c^2}\  MN\ =\sqrt{0.2}\  
\frac {4G}{R_0^2}\ MN \ .
\end{equation} 
(where  $R_0=c/H_0$ is the present horizon). 

We now take  such a result, which was obtained for extremely small
values of $N$, and extrapolate it  up to 
the present horizon $R_0=c/H_0$, i,e., we insert in  formula
(\ref{sigmaf})  the
actual value of $N$, so that the quantity
$MN$ can be identified  with the total visible mass  of the Universe.

Concerning the total visible mass $MN$ of the Universe, one can write
\begin{equation}\label{MN}
MN= \frac 43 \pi\ \rho_{\eff}\ R_0^3\ , 
\end{equation}
with a suitable effective density $\rho_{\eff}$. It is rather  easily
shown (see later) that a quite natural consistency condition of our
model leads to the estimate
\begin{equation}\label{rhoeff}
\rho_{\eff}\simeq \frac 14\,  \frac {3H_0^2}{8\pi G}\simeq  5 \rho_0 \ ,
\end{equation}
where $\rho_0=\Omega_0\, \big({3H_0^2})/\big({8\pi G})$, with 
$\Omega_0\simeq 0.05$, is the actual  density. Inserting
this in (\ref{sigmaf}) one gets $\sigma_f\simeq  0.2\ cH_0$. 

On the other hand, if  a random variable $f$ has  zero mean and
a finite variance $\sigma^2_f$, with great probability its modulus
will take on values very near to its standard deviation $\sigma_f$. 
In such a sense we may say to have found
\begin{equation}\label{forza}
|f|\simeq  0.2 \,  cH_0\ , 
\end{equation}
which perhaps constitutes the main result of the present work.
Namely, in our oversimplified model within the
fractal hypothesis, the force per 
unit mass, i.e., the 
acceleration, exerted by the far matter on a test particle, is found
to have  a value of the order  of $cH_0$, which is the one that
 is needed  in most cases in which 
the presence of a dark matter is advocated. Notice that the assumption
of a uniform, rather than fractal, distribution of matter would lead
instead to $|f|\simeq cH_0 /{\sqrt N}$, i.e., essentially to $f\simeq
0$. Namely,  without the fractal
hypothesis  the Zwicky procedure of neglecting at all the
gravitational contribution of the external matter, would be 
justified.

We can now apply our estimate to the  case of the virial
theorem for a cluster of galaxies. 
To this end, we have  first of all to assume  a property that might in
principle be checked, namely, that the forces at
sufficiently separated  points are uncorrelated. It is then
conceivable  that locally, in some regions, such a random field of
force may form patterns of a
central type, attractive towards a center. This is obviously
equivalent to admit that locally, in such special regions, 
 the external far  matter produces a pressure.

Let us recall that, according to the virial theorem, for a confined
system of $n$ particles one has
\begin{equation}\label{zero}
\overline {\sigma^2_v} = -  \overline \vir\, /n \ ;
\end{equation}
here, $\sigma^2_v=(1/n)\, \sum_i v_i^2$  is the variance of the
velocity distribution of the galaxies  of the cluster, whereas 
$\vir=\sum_i \vett f_i\cdot
\vett x_i$ is called the virial of the forces (per unit mass),  
$\vett x_i$ denoting the position
vector of the $i$--th internal particle  with respect
to the center of mass of the cluster, and $\vett f_i$ the force per
unit mass acting on it, while overline denotes time--average,

So we have to estimate the quantity  $\sum_{i=1}^n  \overline 
{\vett f_i\cdot \vett x_i}$.
 In the  conditions we have assumed, all  terms of such a sum can
 be taken to be equal, so that we just have to estimate one of them.
 We can take 
$\overline {\vett f_i\cdot \vett x_i}\simeq - |\bar f|\ \overline
 {|\vett x_i|}  
$,
with $\overline {|\vett x_i|}\simeq L/4$ where $L$ is
the diameter of the cluster, whereas for estimating the modulus of
the force in the direction  of the center one can make use of 
(\ref{forza}).
So for the velocity variance one gets
\begin{equation}
\label{varianza}
\overline{\sigma_v^2}\simeq 0.2\  \frac{ cH_0 L}4\ , 
\end{equation}
where $L$ is the linear dimension of the cluster.
In the case of Coma, one thus finds a value $\simeq 6 \cdot 10^5
\mathrm{km}^2/\mathrm{sec}^2$, which is very near  to the value $5 \cdot 10^5
\mathrm{km}^2/\mathrm{sec}^2$  reported by Zwicky.

Notice  the linear dependence on $L$ in the formula
 (\ref{varianza}). 
In this connection one may point out  that,
 if the external force were smooth, by a Taylor expansion
about the origin one  would have $\vett f_i$ proportional to
$\vett x_i$, and this would lead to a virial (and thus also a
velocity variance)  proportional
to $L^2$ rather than to $L$. Instead,  the observations 
seem to require a proportionality to $L$.
Apparently, this was first pointed out
by Kazanas and  Mannheim \cite{kazanas}, in a  paper  in which 
some data were reported
 in a range  of $L$ covering  five orders of magnitude (see
 \cite{kazanas}, Fig. 2, 
 page 539).  This
property is also confirmed   by a 
dimensional analysis. Indeed, with the parameters entering the
problem, the square of a velocity can be formed only as $c^2$, or as $cH_0L$
or as  $(H_0L)^2$. But the first term is by far too large, the last term
by far too small, while the term linear in $L$ is indeed about  of the 
correct order of magnitude.

We finally show   how  the estimate (\ref{rhoeff})
for the effective density, namely, $\rho_{\eff}\simeq 5 \rho_0$, 
is obtained. To this end one makes reference to
the metric $g_{\mu\nu}=\eta_{\mu\nu}+h_{\mu\nu}$ with $h_{\mu\nu}$ 
defined by (\ref{campo}), and to the corresponding mean metric
obtained  by averaging with respect to the considered probability density
for the positions of the galaxies.  Denoting the mean by 
$\langle\, .\, \rangle$, the  mean metric is then found to be given by
$$
ds^2= \langle\ {g}_{\mu\nu}\ \rangle\  dx^\mu dx^\nu=
(1-\alpha-3\beta)\, 
c^2 dt^2 -
(1+\alpha+\beta) dl^2
$$
where $dl^2=dx^2+dy^2+dz^2$ and
\begin{equation}
\alpha=\frac {2G}{c^2}\ \langle\ \sum_j \frac {M_j}{|\vett q_j|}\
\rangle \ ,\quad
\beta \raisebox{-0.5ex}{$\stackrel{<}{\sim}$} \frac {4GH_0^2}{3c^4}\  \langle\ \sum_j M_j {|\vett q_j|}\ \rangle\ 
\end{equation}
(the approximation of small velocities was used in the  expression for
 $\beta$). Introducing an effective density $\rho_{\eff}$ such that 
$$
\langle\ \sum \frac {M_j}{|\vett q_j|}\ \rangle
\simeq 4\pi
\rho_{\eff} \ \frac{{R_0}^2}2 \ , \quad
\langle\ \sum M_j {|\vett q_j|}\ \rangle\  \simeq 4\pi \rho_{\eff}
\frac{{R_0}^4}4 \ ,
$$ 
one gets
\begin{equation}\label{aaa}
\alpha\simeq {4\pi G} \rho_{\eff} {{R_0}^2}/{c^2}\ ,\quad
\beta\ <  (2/3) \alpha\ .
\end{equation}
The consistency  condition is now the  requirement that the
expansion rate calculated with the mean metric does
actually coincide with the rate that was introduced into the definition of the
model. This condition  takes the form
\begin{equation}\label{consistenza}
\frac 12\frac {d}{dt} \log \frac {1+\alpha+\beta}{1-\alpha-3\beta}=H_0\ .
\end{equation}
On the other hand from (\ref{aaa}),  using ${\dot R}_0=c$, one gets
$$
\dot \alpha\simeq \frac {8\pi G}{c^2} \ \rho_{\eff}\  R_0c\ , \quad
\dot \beta \simeq \frac 23 \dot\alpha \ .
$$
With  these expressions for $\dot \alpha$ and $\dot \beta$,  the consistency 
condition (\ref{consistenza}) then becomes an algebraic one, which gives
for $\rho_{\eff}$ a  value  that we have rounded off to (\ref{rhoeff}).

\begin{acknowledgments}
We thank George Contopoulos, Christos Efthymiopoulos. Francesco Sylos
 Labini and Rudolf Thun for useful discussions.
\end{acknowledgments}


\begin{thebibliography}{100}
\bibitem{zwi}F. Zwicky,   Helv. Phys. Acta \textbf{5}, 110 (1933);
Astrophys. J. \textbf{86}, 217 (9137).
\bibitem{zeldovich} Ya.B. Zeldovich, 
 I.D.  Novikov,   \textit{Stars and relativity} (Dover, New York 1971).
\bibitem{einstein} A. Einstein,  \textit{The meaning of
  relativity} (Princeton U.P., Princeton 1922).
\bibitem{chandra} S. Chandrasekhar,  Rev. Mod. Phys. \textbf{15}, 1 (1943).
\bibitem{mandel} B. Mandelbrot, \textit{The fractal
 geometry of  nature} (Freeman, New York 1977).
\bibitem{sylos} F. Sylos--Labini, M.  Montuori,
 and L. Pietronero,   Phys. Rep. \textbf{293}, 61 (1998).
\bibitem{kazanas} D. Kazanas,  P.D.  Mannhein, in  \textit{After the
first three minutes}, edited by
S. Holt, C.L.  Bennett, B.V.  Trimble, AIP Conference Proceedings
Vol. 222   (American
Institute of Physics, New York 1991).
\end{thebibliography}
\end{document}